\documentclass{jfm}

\usepackage{graphicx}
\usepackage{newtxtext}
\usepackage{newtxmath}
\usepackage{natbib}
\usepackage{xspace} 
\usepackage{hyperref}
\usepackage{amsmath,amssymb}
\hypersetup{
    colorlinks = true,
    urlcolor   = blue,
    citecolor  = blue,
    linkcolor  = blue,
}

\usepackage{physics}
\usepackage{ragged2e}
\usepackage{caption}
\usepackage{subcaption}
\usepackage{comment}

\title{Linear and nonlinear dynamics of a viscous film on an axisymmetric wavy fiber}

\author{Thota Srinivas\aff{1}
  \corresp{\email{thota.srinivas@icts.res.in}}
 \and Kishor Kumar Sarva\aff{2}}

\affiliation{\aff{1}International Centre for Theoretical Sciences, Bengaluru-560089, Karnataka, India
\aff{2} prajnaLabs, Prajnapramana IP Studios LLP, Telangana, India }

\begin{document}
\maketitle

\begin{abstract}
We study the gravity-driven flow of a viscous film coating a vertical fiber whose radius varies sinusoidally along its length. Using a long-wave model for an axisymmetric film at zero Reynolds number, we compute the steady film that follows the corrugation. We then analyse the stability of this film with Bloch--Floquet theory and follow its nonlinear evolution numerically. A symmetry argument shows that, because gravity makes the film drift, the change in the Rayleigh--Plateau growth rate caused by the corrugation is quadratic in its amplitude $a_s$. Second-order perturbation theory gives the coefficient of this quadratic change in closed form, in agreement with the Bloch spectra. The theory also identifies a Bragg resonance at half the Rayleigh--Plateau wavelength. There, the drift detunes counter-propagating modes, and the growth rate is affected only beyond a critical amplitude proportional to the Bond number $Bo$; without gravity, the correction becomes linear in $a_s$. Time-dependent simulations reveal three nonlinear regimes as $Bo$ decreases. At large $Bo$, a train of beads drifts with stick--slip motion. At intermediate $Bo$, the beads merge into a single large drop that slides about four times faster than beads on a smooth fiber. At small $Bo$, the beads lock to the corrugation. The pinning is transient within the model: the beads release after a time that grows steeply as $Bo$ decreases, while the film between them drains. Requiring the film to rupture before release defines an effective pinning threshold, which we determine for a given rupture thickness.
\end{abstract}


{\bf MSC Codes }  {\it(Optional)} Please enter your MSC Codes here

\section{Introduction}\label{sec:intro}
A liquid film coating a vertical fiber is unstable. Azimuthal surface tension drives a Rayleigh--Plateau (RP) instability that gathers the liquid into beads~\citep{goren1962instability,quere1999fluid}, and gravity makes the beads slide down the fiber. This flow underlies coating processes and the use of fibers for heat and mass transfer, in which the beads enlarge the gas--liquid interface~\citep{weinstein2004coating,grunig2012mass,sadeghpour2020characteristics}. Its dynamics have been studied extensively. Linear stability analyses of the falling film~\citep{solorio1987linear} were followed by long-wave models for thin films~\citep{frenkel1992nonlinear,kalliadasis1994drop}. These models were extended to thick films by retaining the full azimuthal curvature~\citep{kliakhandler2001viscous,craster2006viscous}, and to moderate flow rates and inertia by weighted-residual and integral models~\citep{sisoev2006film,ruyer2008modelling,ruyer2012wavy}. Travelling-wave solutions~\citep{trifonov1992steady}, the convective or absolute nature of the instability~\citep{duprat2007absolute,duprat2009spatial}, and the effects of nozzle geometry~\citep{ji2019dynamics,ji2020modelling}, non-axisymmetric perturbations~\citep{gabbard2021asymmetric} and the full Navier--Stokes problem~\citep{liu2021coating} have also been examined. Depending on fiber radius, film thickness, and flow rate, the beads form a regular train, an irregular train with coalescence, or isolated drops separated by a thin film~\citep{kliakhandler2001viscous,craster2006viscous,ji2019dynamics}.

In almost all of these studies, the fiber is a smooth cylinder. Real fibers, however, are often textured. Textile and functional fibers carry surface structure~\citep{rivero2015nanomaterials}, fibers of non-circular cross-section change the drop dynamics~\citep{xie2021investigation}, and chains of beads or grains are a natural limit of a strongly modulated fiber. \citet{sarva2025dynamics} showed that, for a film flowing down a granular chain, the size of the grains controls whether the film forms regularly spaced drops or coarsens through drop coalescence. For films on planar substrates, wall topography is known to modify the base flow, the linear stability, and the nonlinear waves. This has been shown for inclined wavy planes~\citep{wierschem2003instability}, corrugated surfaces~\citep{trifonov2007stability,trifonov2017nonlinear}, microstructures with sharp corners~\citep{bonart2020stability} and in the context of transfer enhancement~\citep{dietze2019effect}. Reduced models for films on curved substrates provide the framework for such geometries~\citep{roy2002lubrication,roberts2006accurate,wray2017reduced}. How a periodic modulation of the fiber radius interacts with the RP instability, the gravity-driven drift of the beads, and the subsequent bead dynamics has not, to our knowledge, been analysed systematically.

In this paper we address this question with a long-wave model for an axisymmetric film on a fiber whose radius varies sinusoidally, with amplitude $a_s$ and wavelength $\lambda_s$ (\S\,\ref{sec:governing equations}). The model has the structure of the thick-film equation of \citet{craster2006viscous} with a spatially periodic mobility, and depends on the Bond number $Bo$, the fiber-to-film radius ratio $\alpha$, and $(a_s,\lambda_s)$. We compute the steady film that follows the corrugation (\S\,\ref{sec:base-state}). Its linear stability is a periodic-coefficient problem, which we solve with Bloch--Floquet theory (\S\,\ref{sec:linear stability}). We then address three questions.
\begin{enumerate}
\item How strongly does the corrugation change the RP growth rate? We show that the change is quadratic in $a_s$, as a consequence of a symmetry of the problem and of the gravity-driven drift (\S\,\ref{sec:a_s^2 scaling}). We obtain its coefficient in closed form from second-order perturbation theory (\S\,\ref{sec:perturbation}). At the Bragg condition $\lambda_s=\lambda_{RP}/2$, the drift detunes the coupled modes, so that the growth rate changes only beyond a threshold amplitude proportional to $Bo$.
\item What are the nonlinear states? Time-dependent simulations (\S\,\ref{sec:nonlinear}) show a drifting bead train with stick--slip motion, coarsening into a single fast drop, and pinning of the beads to the corrugation.
\item Is the pinning permanent? We find that it is transient within the model. The release time grows steeply as $Bo$ decreases, while the film between the pinned beads drains. A rupture criterion for this film defines an effective pinning threshold.
\end{enumerate}
Section~\ref{sec:conclusions} summarises the results and discusses the limitations of the model.

\section{Governing equation}\label{sec:governing equations}

Using the long-wave approximation with an axisymmetry assumption, we obtain the following evolution equation for the fluid interface $S(z,t)$ on a wavy fiber of mean radius $\alpha$:
\begin{equation}\label{eq:S equation}
    \left.\begin{array}{l}
        8\partial_t\qty(S^2)+\partial_zJ=0,   \\
        J = \qty(\partial_z\kappa +Bo)\,M\qty(S,\, S_s),
    \end{array}\right\}
\end{equation}
where $J$ represents the flux, $M$ represents the lubrication mobility function, and $\kappa$ represents the fluid interface curvature given by
\begin{equation}\label{eq:M and kappa equations}
    \left.\begin{array}{l}
        M\qty(S,\, S_s) = 2S^2\qty(S_s^2-S^2+2S^2\ln\qty(S/S_s))-\qty(S^2-S_s^2)^2,  \\
        \kappa = \partial_{zz}S-S^{-1}.
    \end{array}\right\}
\end{equation}
Here, the fluid interface $S(z,\, t)$ and solid fiber surface $S_s(z)$ are both measured from the axis of symmetry and are defined as
\begin{equation}\label{eq:S and S_s definitions}
    S(z,\, t) = \alpha + \eta(z,\, t),\quad\quad S_s(z) = \alpha+\zeta(z),
\end{equation}
where $\eta$ is the interface distance from the mean fiber radius and $\zeta$ represents the variation of fiber surface about the mean radius $\alpha$ and is taken as:
\begin{equation}
    \zeta(z)=a_s\,\sin\qty({2\,\pi\,z}/{\lambda_s}),
\end{equation}
where $a_s$ and $\lambda_s$ are the amplitude and the wavelength of the solid wavy fiber surface. We briefly outline the derivation of \eqref{eq:S equation} in appendix~\ref{app:derivation}. It is apparent that the governing equation for a viscous film over an axisymmetric wavy fiber, given by~\eqref{eq:S equation}, has the same structure as equation (2.17) of~\cite{craster2006viscous} for a film on a smooth fiber, with only a modification in $S_s$ that captures the wavy nature of the solid fiber surface.

Two structural features are apparent from \eqref{eq:S equation}; \emph{(i) conservation:} it is clear that~\eqref{eq:S equation} conserves the total volume $\int_0^L\,\pi\qty(S^2-S_s^2)\,dz$ with periodic boundary conditions and \emph{(ii) gravity drainage:} the term $Bo$ inside $(\partial_z\kappa+Bo)$ is the gravity term, where $Bo$ is the Bond number defined in~\S\,\ref{sec:scalings}. It breaks the pure-capillary gradient-flow structure and drives drift rather than mere relaxation. In~\eqref{eq:M and kappa equations}, the axial term $\partial_{zz}S$ is stabilizing, and the azimuthal term $-1/S$ is the destabilizing Rayleigh-Plateau (RP) contribution.

\subsection{Scalings and dimensionless parameters}\label{sec:scalings}
We consider a Newtonian liquid of density $\rho$, dynamic viscosity $\mu$ and surface tension $\gamma$ coating a vertical fiber whose radius varies as $R_s(z^*) = R_f + A_s\sin(2\pi z^*/\Lambda_s)$, where $R_f$ is the mean fiber radius and $A_s$ and $\Lambda_s$ are the dimensional amplitude and wavelength of the corrugation; here and below an asterisk denotes a dimensional quantity. Gravity $g$ acts along the positive $z^*$-axis. The liquid volume per corrugation wavelength, $V^*$, sets the reference length
\begin{equation}\label{eq:reference length}
    \bar{R}=\qty(R_f^2+\frac{V^*}{\pi\,\Lambda_s})^{1/2},
\end{equation}
which is the radius of the free surface of a uniform film carrying the same volume on a smooth fiber of radius $R_f$. Radial and axial lengths are both scaled by $\bar{R}$; that is, the long-wave expansion is carried out with $\varepsilon=1$, and the full axial curvature term is retained, as in~\cite{craster2006viscous}. Velocities are scaled by the visco-capillary velocity $\gamma/\mu$, time by $\mu\bar{R}/\gamma$ and pressure by $\gamma/\bar{R}$. With these choices,~\eqref{eq:S equation}--\eqref{eq:M and kappa equations} contain the Bond number
\begin{equation}\label{eq:Bond number}
    Bo=\frac{\rho\, g\,\bar{R}^2}{\gamma}=\qty(\frac{\bar{R}}{\ell_c})^2,\qquad \ell_c=\sqrt{\frac{\gamma}{\rho\, g}},
\end{equation}
which measures gravity drainage relative to capillary pressure gradients, with $\ell_c$ the capillary length. The geometry introduces three further groups:
\begin{equation}\label{eq:geometric parameters}
    \alpha=\frac{R_f}{\bar{R}},\qquad a_s=\frac{A_s}{\bar{R}},\qquad \lambda_s=\frac{\Lambda_s}{\bar{R}}\quad\qty(k_s=2\pi/\lambda_s).
\end{equation}
Here $0<\alpha<1$ is the fiber-to-film radius ratio, so that $1-\alpha$ is the mean film thickness. By~\eqref{eq:reference length}, the dimensionless volume constraint reads $\int_0^{\lambda_s}\pi\qty(S^2-S_s^2)\,dz=\pi\lambda_s\qty(1-\alpha^2)$, and the flat-fiber film has $S\equiv1$. The problem is therefore fully specified by the four parameters $\qty(Bo,\,\alpha,\,a_s,\,\lambda_s)$, summarised in table~\ref{tab:parameters}.

The parameters are constrained by the geometry and the model. The fiber must not pinch off, and the film must cover it everywhere, so $a_s<\alpha$ and $\min_z\qty(S_0-S_s)>0$. The long-wave approximation requires small interfacial and fiber slopes, $\abs{\partial_zS_0}\ll1$ and $a_s\,k_s\ll1$, and axial scales that are long compared with the film radius, $\lambda_s\gg1$. These conditions are well satisfied for $\lambda_s\gtrsim\lambda_{RP}/2$ at the amplitudes considered here, but become marginal for the shortest corrugation wavelengths in table~\ref{tab:parameters}, where $a_s k_s\approx0.25$--$0.3$ and $\lambda_s$ is only a few film radii; results in that range are therefore interpreted qualitatively. In the flat-fiber limit ($a_s=0$), $Bo$ does not affect the growth rate but sets the drift speed (see~\S\,\ref{sec:smooth fiber}). The limit $Bo\to0$ describes thin fibers or films with $\bar{R}\ll\ell_c$, where gravity drainage is weak.

\begin{table}
  \begin{center}
\def~{\hphantom{0}}
  \begin{tabular}{lccc}
      Parameter  & Definition & Range explored (\S\S\,\ref{sec:a_s^2 scaling}, \ref{sec:perturbation}) & Reference value \\[3pt]
       Bond number & $\rho g\bar{R}^2/\gamma$ & $0$--$3$ & $1$ \\
       Fiber-to-film radius ratio & $R_f/\bar{R}$ & -- & $0.6$ \\
       Corrugation amplitude & $A_s/\bar{R}$ & $0.005$--$0.15$ & $0.09$ \\
       Corrugation wavelength & $\Lambda_s/\Lambda_{RP}$ & $0.15$--$3$ & $1$ \\
  \end{tabular}
  \captionsetup{width=\linewidth}
  \caption{\justifying Dimensionless parameters of the problem. Here $\Lambda_{RP}=2\sqrt{2}\,\pi\bar{R}$ is the dimensional Rayleigh--Plateau wavelength of the flat-fiber film (\S\,\ref{sec:smooth fiber}), so that $\Lambda_s/\Lambda_{RP}=\lambda_s/\lambda_{RP}$. The reference values are used in figures~\ref{fig:base-state} and~\ref{fig:bands}.}
  \label{tab:parameters}
  \end{center}
\end{table}

In the next section, we show the steady base-state interface profile $S_0(z)$ for which we perform the stability analysis in the coming sections.

\section{Base state profile}\label{sec:base-state}
\begin{figure}
    \centering
    \includegraphics[width=0.65\linewidth]{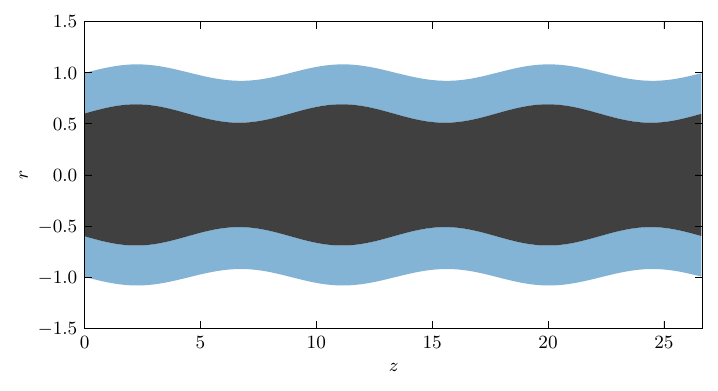}
    \captionsetup{width=\linewidth}
    \caption{\justifying Steady conformal coating $S_0(z)$ (sky-blue) over the wavy fiber $S_s(z)$ (black), corrugation amplitude $a_s=0.09$, fiber-to-film radius ratio $\alpha=0.6$, $Bo=1$ and $\lambda_s=\lambda_{RP}$.}
    \label{fig:base-state}
\end{figure}

At steady-state~\eqref{eq:S equation} becomes $\partial_z\,J=0$ which leads to the following third-order ordinary differential equation for  $S_0(z)$ with $\lambda_s$ periodicity and a constant flux $J_c$:
\begin{equation}\label{eq:base-state}
    \left.\begin{array}{l}
     \qty(\partial_z\kappa_{0}+Bo)\,M\qty(S_0,\,S_s)=J_c,     \\
     \kappa_0 = \partial_{zz}S_0-S_0^{-1}.
    \end{array}\right\}
\end{equation}
The above equation is closed by $\lambda_s$-periodicity and the total volume constraint
\begin{equation}
\int_0^{\lambda_s}\pi\qty(S_0^{2}-S_s^{2})\,dz=\pi\lambda_s\qty(1-\alpha^2),
\qquad S_0(z+\lambda_s)=S_0(z),
\label{eq:volconstraint}
\end{equation}
which fixes $J_c$ and $S_s(z)$ is given by~\eqref{eq:S and S_s definitions}. We obtain $S_0(z)$ by representing it in Fourier space and solving (\ref{eq:base-state}) using Newton's method, continuing from the flat-fiber solution $S_0\equiv1$, $J_c=Bo\,M(1,\,\alpha)$ as the corrugation amplitude $a_s$ increases from zero to a target value. Figure~\ref{fig:base-state} shows the base-state fluid interface $S_0(z)$ (sky blue) over a wavy fiber $S_s(z)$ (black) with $Bo=1$, $\alpha=0.6$, $\lambda_s=\lambda_{RP}$ and $a_s=0.09$. Here, $\lambda_{RP}=2\sqrt{2}\,\pi$ is the most unstable wavelength of the Rayleigh--Plateau (RP) instability for the film on a smooth fiber (\S\,\ref{sec:smooth fiber}).

\section{Linear stability analysis}\label{sec:linear stability}
In this section, we present the linear stability analysis of~\eqref{eq:S equation} about the steady state profile $S_0(z)$. For this, we write the full $S(z,\, t)$ as $S(z,\, t)=S_0(z) + s_1(z,\, t)$ and substitute in~\eqref{eq:S equation}. By retaining only the linear terms in $s_1$, we obtain the following linear perturbation equation for $s_1$:
\begin{equation}\label{eq:linear equation}
    16\,S_0(z)\,\partial_ts_1=-\partial_z\qty[M(S_0,\, S_s)\partial_z\qty{\partial_{zz}s_1+\qty(s_1/S_0^2)}+B(z)\,s_1],
\end{equation}
where $B(z)$ is $\lambda_s$ periodic functions given by
\begin{equation}
    B(z)=\qty(\partial_z\kappa_0+Bo)\eval{\partial_SM}_{(S_0,\, S_s)}=8\qty(\partial_z\kappa_0+Bo)S_0\qty(S_s^2-S_0^2+2S_0^2\ln\qty(S_0/S_s)).
\end{equation}
Since the equation (\ref{eq:linear equation}) is linear with time-independent coefficients, we can expect the solution as $s_1(z,\,t)=e^{\sigma\, t}\phi(z)$. Hence, the equation (\ref{eq:linear equation}) becomes
\begin{equation}\label{eq:linear phi equation}
        16\,\sigma\,S_0\,\phi=-\partial_z\qty[M_0\qty{\partial_{zzz}\phi+\partial_z\qty(\phi/S_0^2)}+B\,\phi],
\end{equation}
a fourth-order, non-self-adjoint eigenvalue problem with $\lambda_s$-periodic coefficients $S_0,\, M_0$ and $B$. Here, $M_0=M(S_0,\, S_s)$. Due to the presence of $\lambda_s$-periodic coefficients in (\ref{eq:linear phi equation}), using Bloch's theorem, one of its solutions can be expressed as
\begin{equation}\label{eq:Bloch form}
\phi(z)=e^{\text{i}\,q\,z}u(z),\quad u(z+\lambda_s)=u(z),\quad
q\in\big(-{k_s}/{2},\, {k_s}/{2}\big],
\end{equation}
where $q$ is the Bloch number and $u(z)$ is the unknown $\lambda_s$-periodic function and thus $\phi(z)$ becomes a quasi-periodic function. Substitution of (\ref{eq:Bloch form}) in (\ref{eq:linear phi equation}) results in
\begin{equation}\label{eq:bloch equation}
    16\,\sigma\,S_0\,u=-\mathscr{D}_q\qty[M_0\qty{\mathscr{D}_{q}^3u+\mathscr{D}_q(u/S_0^2)}+Bu],\quad \text{for }q\in\big(-{k_s}/{2},\, {k_s}/{2}\big],
\end{equation}
where $\mathscr{D}_q=\partial_z+\text{i}\,q$ is the Bloch derivative. Since the coefficients $S_0,\, M_0,\, B$ and the solution $u$ are $\lambda_s$--periodic functions, they can be represented on a Fourier basis. Hence, (\ref{eq:bloch equation}) becomes the following generalised eigenvalue problem after Fourier transform:
\begin{equation}\label{eq:eigenvalue problem}
   \boldsymbol{\mathscr{L}}(q)\hat{\boldsymbol{u}}=\sigma\,\boldsymbol{\mathscr{M}}\hat{\boldsymbol{u}},\quad \text{for }q\in\big(-{k_s}/{2},\,{k_s}/{2}\big],
\end{equation}
where the linear operator $\mathscr{L}(q)$ and the coefficient matrix $\mathscr{M}$ are given by
\begin{gather}
    \boldsymbol{\mathscr{L}}(q)=-\boldsymbol{D}_q\qty[\boldsymbol{M}_0\qty{\boldsymbol{D}_q^3+\boldsymbol{D}_q\boldsymbol{A}}+\boldsymbol{B}],\\
    \boldsymbol{\mathscr{M}}=16\,\boldsymbol{S}_0.
\end{gather}
 Here, $\boldsymbol{D}_q$ is the dense spectral matrix representation for $\partial_q$. Similarly, $\boldsymbol{M}_0,\, \boldsymbol{A},\, \boldsymbol{B}$ and $\boldsymbol{S}_0$ are the diagonal matrices represent the discrete versions of $M_0,\, 1/S_0^2,\, B$ and $S_0$, respectively on a uniform Fourier grid. The film is unstable iff $\max_q\max_j\Re{\sigma_j(q)}>0$ and $-\Im{\sigma_j(q)}/k^*$ gives the drift speed for each Bloch number $q$ and $j\in\qty[1,\, N]$ with $N$ being the number of grid points. Here $k^*=q+n^*k_s$ is the unfolded wavenumber of the mode, with $n^*$ the dominant Fourier harmonic of $u$.
 
\subsection{Smooth fiber limit}\label{sec:smooth fiber}
When the fiber is smooth with $S_s=\alpha+\zeta(z)=\alpha$, the base state interface profile $S_0$ becomes constant ($S_0=1$ with the scaling of~\S\,\ref{sec:scalings}; we retain $S_0$ below for generality). This leads to constant coefficients in (\ref{eq:linear phi equation}). Hence, the solution of the corresponding equation can be expressed as
\begin{equation}\label{eq:normal mode for phi}
    \phi(z)=\hat{\phi}\,e^{\text{i}\,k\,z},
\end{equation}
where $k$ is the wavenumber and $\hat{\phi}$ is the constant complex amplitude. Substitution of (\ref{eq:normal mode for phi}) in (\ref{eq:linear phi equation}) results in the following dispersion relation:
\begin{equation}\label{eq:RP dispersion curve}
    \sigma = \frac{M(S_0,\, \alpha)}{16\,S_0}k^2\qty(\frac{1}{S_0^2}-k^2)-\text{i}\,\frac{Bo\,\partial_SM(S_0,\, \alpha)}{16\,S_0}k,
\end{equation}
where $\partial_SM(S_0,\, \alpha)=8\,S_0\qty(\alpha^2-S_0^2+2\,S_0^2\ln\qty(S_0/\alpha))$. Hence, the most unstable mode occurs at the Rayleigh-Plateau wavenumber $k_{RP}=1\big/\qty(\sqrt{2}\,S_0)$ or the Rayleigh-Plateau wavelength $\lambda_{RP}=2\pi/k_{RP}=2\sqrt{2}\,\pi\, S_0$. The gravity drift is given by $c=-\Im{\sigma}/k=Bo\,\partial_SM(S_0,\, \alpha)/(16\,S_0)$. Thus, on a smooth fiber, the growth rate depends only on $\alpha$ (through $M$), while gravity enters only through the drift speed, which is proportional to $Bo$.

\section{Results}

\subsection{Bloch spectrum}\label{sec:bloch results}
\begin{figure}
     \centering
     \begin{subfigure}[b]{0.49\textwidth}
         \centering
         \includegraphics[width=\textwidth]{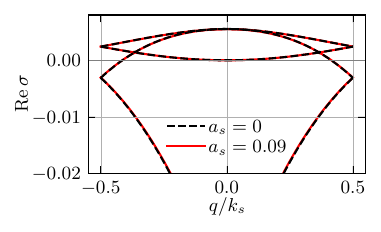}
         \caption{}
         \label{fig:growth rate bands}
     \end{subfigure}
     \hfill
     \begin{subfigure}[b]{0.49\textwidth}
         \centering
         \includegraphics[width=\textwidth]{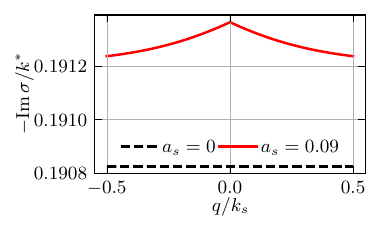}
         \caption{}
         \label{fig:drift bands}
     \end{subfigure}
    \captionsetup{width=\linewidth}
    \caption{\justifying Variation with normalised Bloch number $q/k_s$ of a) growth rate of the first four bands and b) drift speed of the most unstable mode, for $Bo=1$, $\alpha=0.6$ and $\lambda_s=\lambda_{RP}$.}
    \label{fig:bands}
\end{figure}

Figure~\ref{fig:bands} shows the four least stable Bloch bands for $\lambda_s=\lambda_{RP}$. For $a_s=0$ the bands are the smooth-fiber dispersion relation~\eqref{eq:RP dispersion curve} folded into the Brillouin zone at $k=q+n\,k_s$; the Bloch solver reproduces~\eqref{eq:RP dispersion curve} to a relative error of $10^{-11}$. The neutral eigenvalue $\sigma=0$ at $q=0$ corresponds to the volume-conserving shift of the base state and is excluded below. For $a_s=0.09$ the bands are only slightly displaced. The maximum growth rate, attained at $q=0$ ($k^*=k_{RP}$), decreases from $5.527\times10^{-3}$ to $5.499\times10^{-3}$ (by $0.5\,\%$), and the drift speed of the most unstable mode increases from $c=0.1908$ to $0.1912$--$0.1914$.

\subsection{Scaling with the corrugation amplitude}\label{sec:a_s^2 scaling}
\begin{figure}
     \centering
     \begin{subfigure}[b]{0.49\textwidth}
         \centering
         \includegraphics[width=\textwidth]{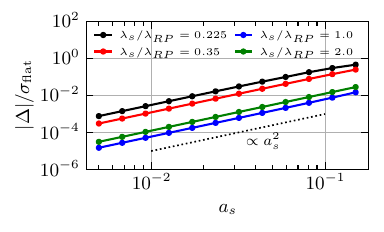}
         \caption{}
         \label{fig:growth rate vs lams}
     \end{subfigure}
     \hfill
     \begin{subfigure}[b]{0.49\textwidth}
         \centering
         \includegraphics[width=\textwidth]{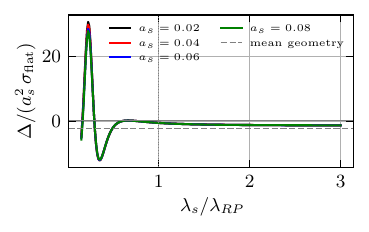}
         \caption{}
         \label{fig:growth rate vs as}
     \end{subfigure}
     \captionsetup{width=\linewidth}
    \caption{\justifying Variation in relative maximum growth rate $\Delta$ with a) $a_s$ at various $\lambda_s/\lambda_{RP}$ and b) $\lambda_s/\lambda_{RP}$ at various $a_s$, for $Bo=1$ and $\alpha=0.6$. Here, in (a), the dotted line shows the reference $a_s^2$ power law, and in (b), the dashed line corresponds to a flat fiber with the same volume and mean-square radii as the wavy one.}
    \label{fig:relative growth with lambda and as}
\end{figure}

Let $\Delta=\max_q\Re\sigma-\max_k\Re\sigma_0$ be the change in the maximum growth rate relative to the smooth fiber. Figure~\ref{fig:relative growth with lambda and as}(a) shows that $\abs{\Delta}\propto a_s^2$ over $0.005\le a_s\le0.15$ at all $\lambda_s$ considered; the fitted exponents lie between $1.993$ and $2.001$. Accordingly, the curves of $\Delta/a_s^2$ for different $a_s$ collapse onto a single function $C(\lambda_s)$ in figure~\ref{fig:relative growth with lambda and as}(b).

This scaling follows from three properties of the problem. First, the Jacobian of~\eqref{eq:base-state} at $a_s=0$ acts on $e^{\text{i}kz}$ as $M_{00}\,\text{i}k(1-k^2)+Bo\,M_{10}$, whose real part $Bo\,M_{10}$ does not vanish; by the implicit function theorem the base state $S_0(z;a_s)$ is therefore a unique analytic function of $a_s$. Second, the translation $z\mapsto z+\lambda_s/2$ maps $\zeta$ to $-\zeta$, so that $S_0(z;-a_s)=S_0(z+\lambda_s/2;\,a_s)$. This translation multiplies a Bloch mode by the constant $e^{\text{i}q\lambda_s/2}$ and leaves $q$ unchanged, so the spectrum at each $q$ is the same for $a_s$ and $-a_s$. Third, because $\Im\sigma_0=-Bo\,M_{10}k/16$ differs for every $k=q+nk_s$, all smooth-fiber eigenvalues at a given $q$ are simple when $Bo>0$, and each depends analytically on $a_s$. An analytic eigenvalue branch whose spectrum is invariant under $a_s\mapsto-a_s$ must be even, so that
\begin{equation}\label{eq:even expansion}
    \sigma(q;a_s)=\sigma_0(q)+a_s^2\,\sigma_2(q)+\order{a_s^4}.
\end{equation}
Numerically, the spectra at $a_s=\pm0.08$ agree to $10^{-12}$. The residual spread of the curves in figure~\ref{fig:relative growth with lambda and as}(b) is the $\order{a_s^4}$ term. The coefficient $C=\Re\sigma_2(k_{RP})$ is obtained in closed form in~\S\,\ref{sec:perturbation}. The dashed line in figure~\ref{fig:relative growth with lambda and as}(b) shows the change produced only by the change in mean geometry, i.e.\ a smooth fiber with the same volume and mean-square radii, for which $C/\sigma_{\rm flat}=-2.31$; the difference between this line and $C$ is the effect of the corrugation itself.

\section{Small-amplitude theory}\label{sec:perturbation}

For $a_s\ll1$ the effect of the corrugation can be computed analytically. We write
\begin{equation}\label{eq:pert expansions}
    S_s=\alpha+a_s\,\zeta,\qquad \zeta=\hat{\zeta}\,e^{\text{i}k_sz}+\text{c.c.},\quad \hat{\zeta}=-\text{i}/2,\qquad
    S_0=1+a_s S_1+a_s^2S_2+\order{a_s^3},
\end{equation}
and denote by $M_{ij}=\partial^{i+j}M/\partial S^i\partial S_s^j$ the derivatives of the mobility evaluated at $(S,S_s)=(1,\,\alpha)$. Let
\begin{equation}\label{eq:sigma0 general}
    \sigma_0(k)=\frac{M_{00}\,k^2\qty(1-k^2)-\text{i}\,Bo\,M_{10}\,k}{16}
\end{equation}
denote the smooth-fiber dispersion relation~\eqref{eq:RP dispersion curve} with $S_0=1$.

\subsection{Base state}\label{sec:pert base}
At $\order{a_s}$,~\eqref{eq:base-state} gives $Bo\qty(M_{10}S_1+M_{01}\zeta)+M_{00}\qty(S_1'''+S_1')=J_1$. The volume constraint~\eqref{eq:volconstraint} gives $\langle S_1\rangle=0$, and therefore $J_1=0$ and
\begin{equation}\label{eq:S1}
    S_1=\hat{s}\,e^{\text{i}k_sz}+\text{c.c.},\qquad \hat{s}=\frac{Bo\,M_{01}\,k_s}{32\,\sigma_0(k_s)}.
\end{equation}
The interface responds to the corrugation through the flat-film operator evaluated at the corrugation wavenumber. In particular, $\hat s\to0$ as $Bo\to0$: without gravity the static interface remains cylindrical. At $\order{a_s^2}$ only the mean of $S_2$ is needed below, and the volume constraint fixes it as $\langle S_2\rangle=\tfrac14-\abs{\hat{s}}^2$.

\subsection{Eigenvalue expansion}\label{sec:pert eigenvalue}
Writing the coefficients of~\eqref{eq:linear phi equation} as $M_0=M_{00}+a_s m_1+a_s^2m_2$, $1/S_0^2=1+a_sA_1+a_s^2A_2$ and $B=Bo\,M_{10}+a_sb_1+a_s^2b_2$, the first-order coefficients carry only the harmonics $e^{\pm\text{i}k_sz}$, with
\begin{equation}\label{eq:first order harmonics}
    \hat m=M_{10}\hat s+M_{01}\hat\zeta,\qquad \hat A=-2\hat s,\qquad
    \hat b=Bo\qty(M_{20}\hat s+M_{11}\hat\zeta)+\text{i}k_s\qty(1-k_s^2)M_{10}\hat s .
\end{equation}
Consider a smooth-fiber mode $\phi=e^{\text{i}kz}$, which belongs to the Bloch number $q\equiv k$ (mod $k_s$). The first-order operator couples it only to $e^{\text{i}(k\pm k_s)z}$. The matrix element from wavenumber $k_b$ to $k_a=k_b+k_s$ is
\begin{equation}\label{eq:P elements}
    P_{ab}=k_ak_b\qty(1-k_b^2)\,\hat m+M_{00}\,k_a^2\,\hat A-\text{i}k_a\,\hat b-16\,\sigma_0(k)\,\hat s,
\end{equation}
and the element to $k_a=k_b-k_s$ is obtained by replacing $(\hat m,\, \hat A,\, \hat b,\, \hat s)$ by their complex conjugates. Since $P$ has no diagonal part, $\sigma_1=0$; this is consistent with the symmetry argument of~\S\,\ref{sec:a_s^2 scaling}. Gravity makes all smooth-fiber eigenvalues at a given $q$ distinct (they differ in $\Im\sigma_0=-Bo\,M_{10}k/16$), so non-degenerate perturbation theory applies and gives
\begin{equation}\label{eq:sigma2}
    \sigma=\sigma_0(k)+a_s^2\,\sigma_2(k)+\order{a_s^4},\qquad
    \sigma_2=\frac{Q}{16}+T_++T_-,\qquad T_\pm=\frac{P_{0,\pm}\,P_{\pm,0}}{256\qty[\sigma_0(k)-\sigma_0(k\pm k_s)]},
\end{equation}
where $P_{\pm,0}$ couples $k\to k\pm k_s$ and $P_{0,\pm}$ couples $k\pm k_s\to k$. The diagonal second-order contribution is
\begin{multline}\label{eq:Q}
    Q=k^2\qty(1-k^2)\langle m_2\rangle+M_{00}k^2\langle A_2\rangle-\text{i}k\langle b_2\rangle
    +k\qty[(k+k_s)\,\hat m^{*}\hat A+(k-k_s)\,\hat m\,\hat A^{*}]-16\,\sigma_0(k)\langle S_2\rangle,
\end{multline}
with
\begin{gather}
    \langle m_2\rangle=M_{10}\langle S_2\rangle+M_{20}\abs{\hat s}^2-M_{11}\,\Im\hat s+\tfrac14M_{02},\qquad
    \langle A_2\rangle=-2\langle S_2\rangle+6\abs{\hat s}^2,\\
    \langle b_2\rangle=Bo\qty(M_{20}\langle S_2\rangle+M_{30}\abs{\hat s}^2-M_{21}\,\Im\hat s+\tfrac14M_{12})-k_s\qty(1-k_s^2)M_{11}\,\Re\hat s .
\end{gather}
The terms $T_\pm$ in~\eqref{eq:sigma2} describe the scattering of the mode $k$ into $k\pm k_s$ by the corrugation and back. The term $Q/16$ collects the change in the mean film geometry and the self-interaction of the first-order base-state correction. Evaluating~\eqref{eq:sigma2} at the most unstable smooth-fiber wavenumber gives the coefficient of figure~\ref{fig:relative growth with lambda and as}(b) in closed form:
\begin{equation}\label{eq:C analytic}
    \Delta=C\,a_s^2+\order{a_s^4},\qquad C\qty(\lambda_s,Bo,\alpha)=\Re\,\sigma_2\qty(k_{RP}),\quad k_{RP}=1/\sqrt2 .
\end{equation}
The imaginary part of $\sigma_2$ gives the $\order{a_s^2}$ correction to the drift speed. Figure~\ref{fig:C analytic} compares~\eqref{eq:C analytic} with the Bloch eigenvalues; the two agree to within a relative error of $2\times10^{-6}$ when $\sigma_2$ is extracted from the numerics by Richardson extrapolation in $a_s$. Decomposing $C$ into its three terms shows that, for $Bo=1$ and $\lambda_s\gtrsim\lambda_{RP}/2$, the forward-scattering term $T_+$ (into $k_{RP}+k_s$) is stabilizing and dominates the mean-geometry term $Q/16$. The large positive and negative values of $C$ for $\lambda_s\lesssim0.4\,\lambda_{RP}$ come from the backward-scattering term $T_-$, whose sign is set by the phase of $P_{0,-}P_{-,0}/[\sigma_0(k_{RP})-\sigma_0(k_{RP}-k_s)]$.

\subsection{Bragg resonance and the role of gravity}\label{sec:bragg}
\begin{figure}
     \centering
     \begin{subfigure}[b]{0.49\textwidth}
         \centering
         \includegraphics[width=\textwidth]{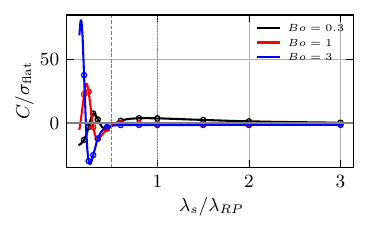}
         \caption{}
         \label{fig:C analytic}
     \end{subfigure}
     \hfill
     \begin{subfigure}[b]{0.49\textwidth}
         \centering
         \includegraphics[width=\textwidth]{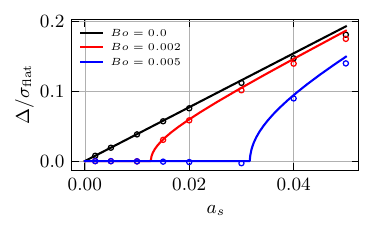}
         \caption{}
         \label{fig:Bragg}
     \end{subfigure}
     \captionsetup{width=\linewidth}
    \caption{\justifying (a) Second-order coefficient $C/\sigma_{\rm flat}$ from~\eqref{eq:C analytic} (lines) and from the Bloch eigenvalues with $a_s=0.005$ (symbols), for $\alpha=0.6$ and several $Bo$. The dashed line marks the Bragg condition $\lambda_s=\lambda_{RP}/2$. (b) Change of the maximum growth rate at $\lambda_s=\lambda_{RP}/2$ for small $Bo$: two-mode prediction~\eqref{eq:Bragg} (lines) and Bloch eigenvalues (symbols).}
    \label{fig:perturbation}
\end{figure}

Expansion~\eqref{eq:sigma2} fails when a denominator of $T_\pm$, $\sigma_0(k)-\sigma_0(k\pm k_s)$, becomes small. For the most unstable mode, this occurs at the Bragg condition $k_s=2k_{RP}$ ($\lambda_s=\lambda_{RP}/2$), where $k_{RP}$ and $k_{RP}-k_s=-k_{RP}$ have equal growth rates and differ only through the drift, $\sigma_0(k_{RP})-\sigma_0(-k_{RP})=-\text{i}\,Bo\,M_{10}k_{RP}/8$. Retaining these two modes exactly gives a $2\times2$ eigenvalue problem. For $Bo\to0$, where $\hat s=\order{Bo}$ and $P_{0,\,-}P_{-,\,0}\to M_{01}^2/64$, its solution is
\begin{equation}\label{eq:Bragg}
    \max\Re\sigma=\frac{M_{00}}{64}+\frac{\abs{M_{01}}}{128}\sqrt{a_s^2-a_c^2}\;\;(a_s>a_c),\qquad
    a_c=\frac{8\,Bo\,M_{10}\,k_{RP}}{\abs{M_{01}}},
\end{equation}
while for $a_s<a_c$ the growth rate is unchanged to this order, and the correction is $\order{a_s^2}$. Hence the $a_s^2$ law of~\S\,\ref{sec:a_s^2 scaling} holds only for $a_s\ll a_c\propto Bo$. In the absence of gravity ($Bo=0$) the corrugation increases the growth rate linearly, $\Delta=\abs{M_{01}}a_s/128$. Physically, the drift detunes the counter-propagating modes $\pm k_{RP}$ by $2ck_{RP}$, and the corrugation can lock them into a faster-growing standing pattern only when the coupling it provides exceeds this detuning. The threshold~\eqref{eq:Bragg} has the form of an exceptional point of the non-normal $2\times2$ problem. Figure~\ref{fig:Bragg} confirms~\eqref{eq:Bragg} against the Bloch solver. A similar loss of validity of~\eqref{eq:sigma2} occurs for long corrugations ($k_s\to0$) at small $Bo$, where $k_{RP}\pm k_s$ become nearly degenerate with $k_{RP}$; there the corrugation acts as a slowly varying film thickness and the relevant threshold scales as $Bo\,k_s$.

\section{Nonlinear dynamics}\label{sec:nonlinear}

\subsection{Numerical method}\label{sec:nonlinear method}
We integrate~\eqref{eq:S equation} in time on a periodic domain containing $P$ corrugation wavelengths, $0\le z<P\lambda_s$. The unknown is $u=S^2$, which satisfies
\begin{equation}\label{eq:evolution u}
    \partial_t u=-\tfrac18\,\partial_zJ,\qquad J=\qty(\partial_z\kappa+Bo)\,M\qty(\sqrt{u},\,S_s),
\end{equation}
discretised with a Fourier pseudo-spectral method on $N$ equispaced points. Because the spectral derivative of $J$ has zero mean, the liquid volume $\int\pi(S^2-S_s^2)\,dz$ is conserved to round-off; the relative change over all runs reported here is below $10^{-14}$. The resulting system is stiff because of the fourth-order capillary term, and we advance it with a variable-order, variable-step backward-differentiation scheme using the exact Jacobian,
\begin{equation}\label{eq:jacobian}
    \frac{\partial\dot{\boldsymbol u}}{\partial\boldsymbol u}=-\frac18\,\boldsymbol D\qty[\boldsymbol M\qty(\boldsymbol D^3+\boldsymbol D\boldsymbol A)+\operatorname{diag}\qty(G\,\partial_SM)]\operatorname{diag}\qty(\frac{1}{2S}),\qquad G=\partial_z\kappa+Bo,
\end{equation}
where $\boldsymbol D$ is the Fourier differentiation matrix and $\boldsymbol M$, $\boldsymbol A$ are the diagonal matrices of $M$ and $1/S^2$. Each simulation starts from the base state $S_0(z)$ of~\S\,\ref{sec:base-state} with a volume-preserving perturbation of amplitude $10^{-3}$, composed of the first $4P$ Fourier modes of the domain with random phases. The position of the bead pattern, $x(t)$, is obtained from the phase of the dominant Fourier mode of $S-S_0$, and the mean speed $\bar v$ from the second half of each run.

\emph{Validation.} Starting from a Bloch eigenmode of~\S\,\ref{sec:linear stability} with amplitude $10^{-6}$, the computed growth rate and drift agree with the eigenvalue $\sigma$ to a relative error of $4\times10^{-5}$, decreasing to $3\times10^{-6}$ when the time-integration error is reduced. This was checked for $a_s=0$ and $0.09$, for $\lambda_s/\lambda_{RP}=1$, $0.5$ and $0.225$, and for Bloch numbers $q=0$, $k_s/2$ and $2\pi/(4\lambda_s)$. With $P=4$, saturated states computed with $N=257$ and $N=385$ agree to $4\times10^{-5}$ in amplitude. The thin films that form between pinned beads (\S\,\ref{sec:pinning}) are resolved with $N=385$--$513$ down to a thickness of about $10^{-3}$. Below this thickness, the computed drainage depends on $N$. We therefore use $N=513$ near the pinning threshold, and report only quantities that change by less than $3\,\%$ between $N=385$ and $N=513$.

\subsection{Saturation and drift on smooth and wavy fibers}\label{sec:nonlinear regimes}
We fix $\alpha=0.6$, $\lambda_s=\lambda_{RP}$ and $P=4$, so that the most unstable linear mode places one bead per corrugation wavelength, and vary $Bo$. Table~\ref{tab:nonlinear} summarises the saturated states.

On the smooth fiber, the Rayleigh--Plateau instability saturates into a train of four travelling beads in all cases. The Bond number controls the bead amplitude. For $Bo=1$ the saturated state is a small-amplitude travelling wave (interface half-amplitude $0.054$) that drifts essentially at the linear speed $c=Bo\,\partial_SM/16$. As $Bo$ decreases, the beads become large and are connected by a thin film, and their speed falls increasingly below $c$: $\bar v/c=0.99$, $0.94$, $0.76$ and $0.45$ for $Bo=1$, $0.3$, $0.1$ and $0.03$.

\begin{figure}
    \centering
    \includegraphics[width=\linewidth]{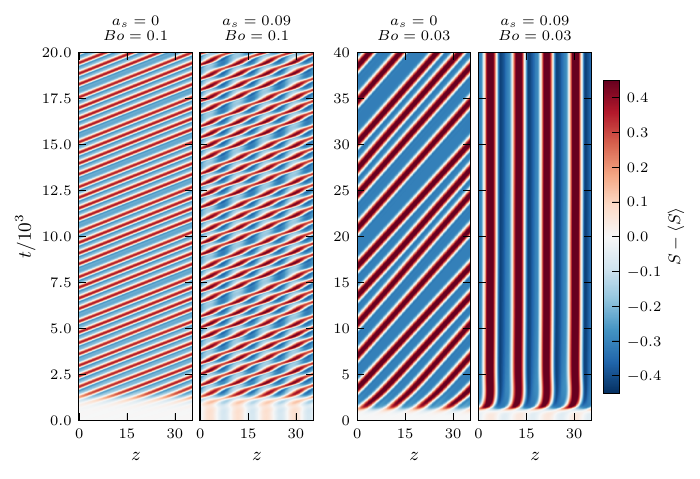}
    \captionsetup{width=\linewidth}
    \caption{\justifying Space--time diagrams of $S-\langle S\rangle$ on smooth ($a_s=0$) and wavy ($a_s=0.09$) fibers for $Bo=0.1$ (left pair, $t\le2\times10^4$) and $Bo=0.03$ (right pair, $t\le4\times10^4$), with $\alpha=0.6$, $\lambda_s=\lambda_{RP}$ and four corrugation wavelengths in the domain. Tilted stripes are drifting beads; vertical stripes are pinned beads.}
    \label{fig:spacetime}
\end{figure}

\begin{figure}
    \centering
    \includegraphics[width=\linewidth]{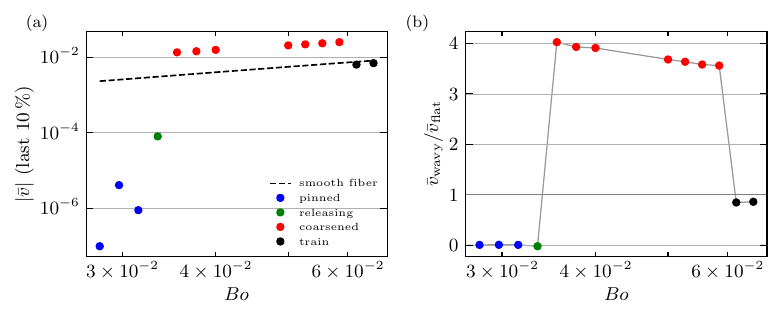}
    \captionsetup{width=\linewidth}
    \caption{\justifying Final state and mean speed over the last $10\,\%$ of runs of duration $t=4\times10^4$, for $a_s=0.09$, $\alpha=0.6$, $\lambda_s=\lambda_{RP}$, $P=4$ and $N=257$. (a) Absolute speed, with the smooth-fiber value shown dashed. (b) Ratio of the wavy-fiber speed to the smooth-fiber speed. Symbols: pinned (blue), releasing during the run (green), coarsened into a single drop (red), drifting bead train (black).}
    \label{fig:sweep}
\end{figure}

On the wavy fiber with $a_s=0.09$, three regimes appear as $Bo$ decreases (figures~\ref{fig:spacetime}--\ref{fig:sweep}).
\begin{enumerate}
\item For $Bo\ge0.061$ the beads form a drifting train whose speed is modulated as they pass over each corrugation (stick--slip motion). At $Bo=0.1$, for example, the instantaneous speed oscillates between $0.0107$ and $0.0171$ about a mean of $0.0138$, compared with the uniform $0.0146$ on the smooth fiber. Over $0.061\le Bo\le0.1$ the mean speed is $84$--$94\,\%$ of the smooth-fiber value.
\item For $0.035\le Bo\le0.058$ the four beads first form a train that is held back by the corrugation. After a time of $7\times10^3$--$1.8\times10^4$ they merge and coarsen into a single large drop per domain (maximum radius $S\approx2.1$--$2.3$). This drop slides over the corrugation $3.6$--$4.0$ times faster than the smooth-fiber bead train. The transition from this regime to the drifting train is abrupt and occurs between $Bo=0.058$ and $0.061$. 
\item For $Bo\lesssim0.034$ the beads are pinned for times longer than $2.5\times10^4$ before they release and coarsen, or remain pinned while the film between them drains (\S\,\ref{sec:pinning}).
\end{enumerate}
The same regimes were obtained for four different random initial perturbations at $Bo=0.03$ and $0.045$.

\subsection{Pinning, release and film drainage}\label{sec:pinning}
\begin{figure}
    \centering
    \includegraphics[width=\linewidth]{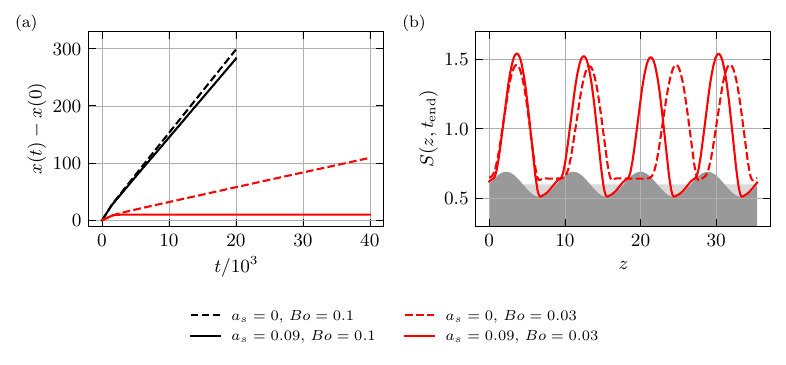}
    \captionsetup{width=\linewidth}
    \caption{\justifying (a) Position of the bead pattern, $x(t)-x(0)$, for the four cases of figure~\ref{fig:spacetime}. The mean speeds are $1.5\times10^{-2}$, $1.4\times10^{-2}$, $2.6\times10^{-3}$ and $3\times10^{-6}$. (b) Interface profiles at $t=4\times10^4$ for $Bo=0.03$ on the smooth and wavy fibers; the shaded region is the wavy fiber.}
    \label{fig:pinning}
\end{figure}

In the pinned state, each bead sits on the downstream flank of a corrugation crest: its maximum lies at $z/\lambda_s\approx0.40$ (mod 1), whereas the crest is at $z/\lambda_s=0.25$ (figure~\ref{fig:pinning}b). The beads creep forward only at speeds of $10^{-7}$--$10^{-5}$, more than three orders of magnitude below the smooth-fiber value, and the film between them drains near the corrugation trough ($z/\lambda_s\approx0.78$). The pinned state is therefore not steady: the film keeps thinning while the beads remain in place.

\begin{figure}
    \centering
    \includegraphics[width=\linewidth]{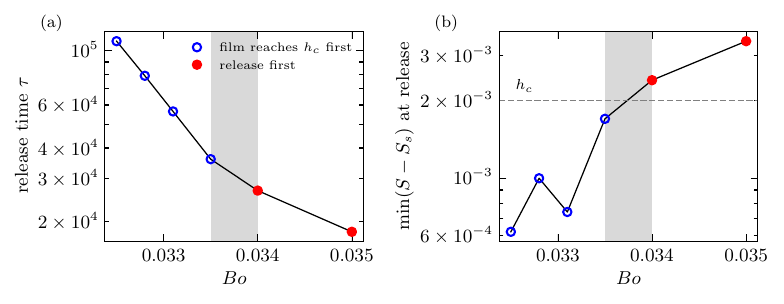}
    \captionsetup{width=\linewidth}
    \caption{\justifying Release of transiently pinned beads for $a_s=0.09$ (table~\ref{tab:release}). (a) Release time $\tau$ against $Bo$. (b) Minimum film thickness at release; the dashed line is the rupture thickness $h_c=2\times10^{-3}$. Open symbols: the film reaches $h_c$ before release; filled symbols: the beads release first. The shaded band marks the pinning threshold $0.0335<Bo_c<0.034$ for this $h_c$.}
    \label{fig:release}
\end{figure}

For $Bo$ just below the coarsening regime, the pinning is transient. After a time $\tau$ the beads release, merge, and coarsen as in regime~(ii). Table~\ref{tab:release} and figure~\ref{fig:release}(a) show that $\tau$ grows steeply as $Bo$ decreases, from $1.8\times10^4$ at $Bo=0.035$ to $1.1\times10^5$ at $Bo=0.0325$. The release time is insensitive to resolution: $N=385$ and $N=513$ give values within $3\,\%$, although the film thickness at release differs by up to a factor of two between the two resolutions. Release is therefore controlled by the bead dynamics rather than by the detailed structure of the draining film.

The variation of $\tau$ with $Bo$ is not described by the inverse-square-root law $\tau\propto(Bo-Bo_c)^{-1/2}$ expected if the pinned state disappeared at a saddle-node bifurcation: such a fit deviates from the data by more than a factor of three. Nor is it exponential in $Bo$, which deviates by up to $22\,\%$. Power laws $\tau\propto(Bo-Bo_c)^{-n}$ with $n=2$--$3$ fit the six values to within $6\,\%$, with $Bo_c\approx0.029$--$0.031$. The data thus indicate that the release time diverges at a Bond number of about $0.03$, but they do not determine the exponent.

Whether the beads remain pinned indefinitely is decided by a competition between release and drainage. As $Bo$ decreases, the film between the beads is thinner at the moment of release (figure~\ref{fig:release}b). In a real liquid, a sufficiently thin film ruptures, or dewets, at a thickness $h_c$ set by intermolecular forces that are absent from~\eqref{eq:S equation}. If the film reaches $h_c$ before the beads release, the beads remain pinned. Taking $h_c=2\times10^{-3}$, which is resolved by the present computations and corresponds to $2\,\mu$m for $\bar R=1$\,mm, we find:
\begin{itemize}
\item At $Bo=0.0335$ the film reaches $h_c$ at $t\approx3.4\times10^4$, before the release at $3.6\times10^4$, so the beads stay pinned.
\item At $Bo=0.034$ the beads release at $t\approx2.7\times10^4$ while the film is still $2.4\times10^{-3}$ thick.
\end{itemize}
For $a_s=0.09$ the pinning threshold is therefore $0.0335<Bo_c<0.034$ for this value of $h_c$. The threshold depends on $h_c$ and moves to smaller $Bo$ as $h_c$ decreases. A disjoining pressure in~\eqref{eq:S equation} would replace this criterion with a prediction of the long-time state, and is left for future work.

\begin{table}
  \begin{center}
  \begin{tabular}{ccccccc}
      & \multicolumn{2}{c}{half-amplitude of $S$} & \multicolumn{3}{c}{mean speed $\bar v$} & \\
      $Bo$ & $a_s=0$ & $a_s=0.09$ & $c$ (linear) & $a_s=0$ & $a_s=0.09$ & state ($a_s=0.09$) \\[3pt]
      $1$    & $0.054$ & $0.095$ & $0.191$  & $0.190$  & $0.190$ & drifting train \\
      $0.3$  & $0.157$ & $0.189$ & $0.0572$ & $0.0540$ & $0.0543$ & drifting train \\
      $0.1$  & $0.314$ & $0.398$ & $0.0191$ & $0.0146$ & $0.0138$ & stick--slip train \\
      $0.03$ & $0.415$ & $0.496$ & $0.0057$ & $0.0026$ & $3\times10^{-6}$ & pinned \\
  \end{tabular}
  \caption{\justifying Saturated states for $\alpha=0.6$, $\lambda_s=\lambda_{RP}$, $P=4$ and $N=257$. The interface half-amplitude is $(\max S-\min S)/2$ at saturation, and the linear drift speed is $c=Bo\,\partial_SM(1,\alpha)/16$.}
  \label{tab:nonlinear}
  \end{center}
\end{table}

\begin{table}
  \begin{center}
  \begin{tabular}{cccc}
      $Bo$ & release time $\tau$ & $\min(S-S_s)$ at release & state for $h_c=2\times10^{-3}$ \\[3pt]
      $0.0325$ & $1.09\times10^5$ & $6.2\times10^{-4}$ & pinned \\
      $0.0328$ & $7.90\times10^4$ & $1.0\times10^{-3}$ & pinned \\
      $0.0331$ & $5.65\times10^4$ & $7.4\times10^{-4}$ & pinned \\
      $0.0335$ & $3.60\times10^4$ & $1.7\times10^{-3}$ & pinned \\
      $0.0340$ & $2.68\times10^4$ & $2.4\times10^{-3}$ & released \\
      $0.0350$ & $1.82\times10^4$ & $3.4\times10^{-3}$ & released \\
  \end{tabular}
  \caption{\justifying Release of transiently pinned beads for $a_s=0.09$, $\alpha=0.6$, $\lambda_s=\lambda_{RP}$ and $P=4$. Release is defined as the first time at which $\max S>1.55$; the pinned beads have $\max S\approx1.52$--$1.53$. Values are for $N=513$, except $Bo=0.0331$ ($N=385$). Release times differ by less than $3\,\%$ between $N=385$ and $N=513$. Film thicknesses below about $10^{-3}$ are not converged.}
  \label{tab:release}
  \end{center}
\end{table}

\section{Conclusions}\label{sec:conclusions}

We have studied a viscous film coating a vertical fiber whose radius varies sinusoidally, using a long-wave model for an axisymmetric film at zero Reynolds number. The results concern the steady film, its linear stability, and its nonlinear evolution.

\emph{Base state and linear stability.} The steady film follows the corrugation with a slightly reduced amplitude and a small phase lag. It is unstable to the RP mechanism, and its linear stability is described by Bloch bands. For the reference case ($Bo=1$, $\alpha=0.6$, $\lambda_s=\lambda_{RP}$, $a_s=0.09$), the corrugation reduces the maximum growth rate by only $0.5\,\%$ and increases the drift speed slightly. In general, the change in the maximum growth rate is $\Delta=C(\lambda_s,Bo,\alpha)\,a_s^2+\order{a_s^4}$. The absence of a linear term follows from two facts: a half-period translation maps $a_s$ to $-a_s$, and the gravity-driven drift makes all smooth-fiber modes at a given Bloch number distinct. Second-order perturbation theory gives $C$ in closed form, in agreement with the Bloch spectra to $2\times10^{-6}$. For $Bo=1$ the coefficient is small for $\lambda_s\gtrsim\lambda_{RP}/2$, where scattering into the shorter mode $k_{RP}+k_s$ is weakly stabilised. It is large and changes sign for shorter corrugations, where scattering into $k_{RP}-k_s$ dominates. Its sign and magnitude also depend on $Bo$: at $Bo=0.3$ the corrugation is destabilising at $\lambda_s=\lambda_{RP}$.

\emph{Bragg resonance and gravity.} At $\lambda_s=\lambda_{RP}/2$, the modes $\pm k_{RP}$ have equal growth rates and are coupled at first order by the corrugation; only the drift separates them. A two-mode analysis shows that the growth rate is unchanged for $a_s<a_c=8\,Bo\,M_{10}k_{RP}/\abs{M_{01}}$ and increases as $\sqrt{a_s^2-a_c^2}$ beyond this threshold, which is an exceptional point. Without gravity, the corrugation increases the growth rate linearly in $a_s$. The drift of the film therefore protects the RP instability from first-order effects of the substrate.

\emph{Nonlinear dynamics.} For $a_s=0.09$ and four corrugation wavelengths per domain, the simulations show three regimes as $Bo$ decreases.
\begin{enumerate}
\item For $Bo\ge0.061$ the beads form a drifting train whose speed is modulated by the corrugation (stick--slip motion), with a mean speed $84$--$94\,\%$ of that on a smooth fiber.
\item For $0.035\le Bo\le0.058$ the beads merge into a single drop that slides $3.6$--$4$ times faster than the smooth-fiber train.
\item For $Bo\lesssim0.034$, the beads lock to the corrugation, each sitting on the downstream flank of a crest.
\end{enumerate}
The pinning is not permanent within the model. The beads release after a time that is independent of resolution and grows steeply as $Bo$ decreases, from $1.8\times10^4$ at $Bo=0.035$ to $1.1\times10^5$ at $Bo=0.0325$. The growth is consistent with a divergence near $Bo\approx0.03$ but not with a saddle-node scaling. Meanwhile, the film between the pinned beads drains. If this film ruptures at a thickness $h_c$, the beads remain pinned whenever rupture precedes release. For $h_c=2\times10^{-3}$, this gives the threshold $0.0335<Bo_c<0.034$.

\emph{Limitations and outlook.} The model is a long-wave approximation in which the full curvature is retained heuristically. Its quantitative predictions are therefore most reliable for corrugations that are long and gentle compared with the film radius ($\lambda_s\gtrsim\lambda_{RP}/2$ and $a_sk_s\ll1$), whereas the large values of $C$ found for short corrugations should be checked against Stokes or Navier--Stokes computations. Inertia, which matters at higher flow rates~\citep{ruyer2008modelling,ruyer2012wavy}, and non-axisymmetric perturbations~\citep{gabbard2021asymmetric} were not considered. The nonlinear simulations used a periodic domain of four corrugation wavelengths, which limits the size of the coalesced drops. The long-time fate of pinned beads depends on intermolecular forces in the draining film. Including a disjoining pressure would replace the rupture criterion used here with a prediction of the final state. Natural extensions are a weakly nonlinear description of bead locking to the corrugation, larger domains and other values of $a_s$ to construct a regime diagram in the $(Bo,a_s)$ plane, and experiments on fibers with controlled corrugation or on granular chains~\citep{sarva2025dynamics}.

\appendix
\section{Axisymmetric long-wave flow on a wavy fiber}\label{app:derivation}
In this section, we present the derivation of~\eqref{eq:S equation} by closely following~\cite{wray2017reduced}. 
\subsection{Geometry and coordinate system}
The reference surface $\mathcal{P}$ is a cylinder with radius $\alpha$
\begin{equation}
    \boldsymbol{P}(z,\, \theta)=\qty(\alpha\cos\theta,\, \alpha\sin\theta,\, z),
\end{equation}
with $y$ measuring distance along an outward unit normal radial vector $\boldsymbol{e}_3$ from the reference surface $\mathcal{P}$. Hence, any point in the fluid is at a distance $\alpha+y$ from the axis. By considering $\qty(x_1,\, x_2)=\qty(z,\, \theta)$, the metric scale factors for $\mathcal{P}$ are given by
\begin{equation}\label{eq:scale factors}
    \qty(m_z,\, m_\theta)=\qty(1,\, \alpha).
\end{equation}
The principal curvatures of the reference surface $\kappa_i$ are defined by $\partial \boldsymbol{e}_3/\partial x_i=-m_i\,\kappa_i\,\boldsymbol{e}_i$. Hence, for the cylindrical surface $\mathcal{P}$
\begin{equation}\label{eq:ref. solid curvatures}
    \qty(\kappa_z,\, \kappa_\theta)=\qty(0,\, -1/\alpha).
\end{equation}
The scale factors within the fluid are given by $h_i=m_i\qty(1-\kappa_i\,y)$, that is
\begin{equation}
    \qty(h_z,\, h_\theta)=\qty(1,\, \alpha+y),
\end{equation}
and at the fluid interface $y=\eta(z,\, t)$, these scale factors are given by $\qty(\Tilde{h}_z,\, \Tilde{h}_\theta)=\qty(1,\, \alpha+\eta)$. The solid surface is kept at $y=\zeta(z)$, with
\begin{equation}
    \zeta(z)=a_s\,\sin\qty(k_s\, z), \quad\quad k_s={2\,\pi}/{\lambda_s},
\end{equation}
where $a_s$ and $\lambda_s$ are the amplitude and the wavelength ($k_s$ as the corresponding wavenumber) of the solid surface. We write throughout 
\begin{equation}
    S(z,\, t) = \alpha + \eta(z,\, t),\quad\quad  S_s(z) = \alpha + \zeta(z),
\end{equation}
for the radial positions of the free fluid interface and the solid surface, respectively. Thus $\alpha$ represents the mean solid surface radius and $S>S_s$ in the fluid.

\subsection{Interfacial curvature}
The curvature of the free surface in this geometry is given by~\citep{wray2017reduced},
\begin{equation}
    \kappa=\frac{1}{\Tilde{h}_1\,\Tilde{h}_2}\qty[\pdv{x_1}\qty(\frac{\Tilde{h}_2^2\,\partial_{x_1}\eta}{\mathscr{A}})+\pdv{x_2}\qty(\frac{\Tilde{h}_1^2\,\partial_{x_2}\eta}{\mathscr{A}})]+\frac{1}{\mathscr{A}}\qty[\qty(\Tilde{h}_1^2+\qty(\partial_{x_1}\eta)^2)\frac{m_2\,\kappa_2}{\Tilde{h}_1}+\qty(\Tilde{h}_2^2+\qty(\partial_{x_2}\eta)^2)\frac{m_1\,\kappa_1}{\Tilde{h}_2}],
\end{equation}
with $\mathscr{A}=\sqrt{\Tilde{h}_1^2\,\Tilde{h}_2^2+\Tilde{h}_2^2\,\qty(\partial_{x_1}\eta)^2+\Tilde{h}_1^2\,\qty(\partial_{x_2}\eta)^2}$. Thus, for the present axisymmetric case $\partial_{x_2}\equiv\partial_\theta\equiv0$ with $\qty(x_1,\, x_2)=\qty(z,\, \theta)$, $\Tilde{h}_1=1$, $\Tilde{h}_2=\alpha+\eta=S$, and from (\ref{eq:scale factors})--(\ref{eq:ref. solid curvatures}): $m_z\,\kappa_z=0$, and $m_\theta\, \kappa_\theta=-1$ such that
\begin{equation}
    \mathscr{A}=S\sqrt{1+\qty(\partial_z\eta)^2},\quad\quad \kappa = \frac{1}{S}\pdv{z}\qty(\frac{S\,\partial_z\eta}{\sqrt{1+\qty(\partial_z\eta)^2}})-\frac{\sqrt{1+\qty(\partial_z\eta)^2}}{S}.
\end{equation}
Since, $\partial_z\eta=\partial_zS$, the above equation can be re-written as
\begin{equation}
    \kappa = \frac{\partial_{zz}S}{\qty(1+\qty(\partial_zS)^2)^{3/2}}-\frac{1}{S\qty(1+\qty(\partial_zS)^2)^{1/2}}.
\end{equation}
Therefore, with long-wave approximation $\abs{\partial_zS}\ll 1$, the above equation can be approximated as
\begin{equation}
    \kappa = \pdv[2]{S}{z}-\frac{1}{S},
\end{equation}
where $\partial_{zz}S$ and $1/S$ terms represent the axial stabilizing and azimuthal destabilizing terms, respectively.

\subsection{Long-wave Stokes problem}
At $Re=0$, with $\varepsilon=1$, the $z-$component of momentum equation from~\cite{wray2017reduced} equation (3.9) to the leading order
\begin{equation}\label{eq:z-momentum at order e}
    \pdv{p}{z}-Bo=\frac{1}{\alpha+y}\pdv{y}\qty[\qty(\alpha+y)\,\pdv{u_z}{y}],
\end{equation}
written in the scalings of~\S\,\ref{sec:scalings}, where gravity acts along the positive $z$-axis, i.e.\ along $x_1$ in $(x_1,\,x_2,\,y)$, and $u_z$ is the fluid velocity along the $z$-axis. Choosing the visco-capillary velocity scale $\gamma/\mu$ sets the capillary number $Ca=\mu U/\gamma$ to unity, so that the gravity term appears with the coefficient $Bo$ in~\eqref{eq:Bond number}. To leading order, the radial momentum equation gives $\partial p/\partial y = 0$. Hence, throughout the fluid layer, the pressure is $p(z,\, t)=-\kappa$. Substitute the expression for the pressure in~\eqref{eq:z-momentum at order e} and the resulting equation for $u_z(z,\, y,\, t)$ is given by
\begin{equation}\label{eq:uz equation wavy}
    -\qty(\pdv{\kappa}{z}+Bo)=\frac{1}{\alpha+y}\,\pdv{y}\qty[\qty(\alpha+y)\,\pdv{u_z}{y}].
\end{equation}
The above equation is solved by subjecting it to the following no-slip condition on the solid surface and zero tangential stress at the free surface:
\begin{gather}\label{eq:bcs for uz}
    \eval{u_z}_{y=\zeta}=0,\quad\quad \eval{\pdv{u_z}{y}}_{y=\eta}=0.
\end{gather}
The kinematic boundary condition for this case is given by
\begin{equation}\label{eq:kinematic bc wavy}
   \qty(\alpha+\eta)\pdv{\eta}{t}+\pdv{z}\int_\zeta^\eta\,\qty(\alpha+y)\,u_z(z,\, y,\, t)\,dy=0. 
\end{equation}

\subsection{Velocity field}
Integration of~\eqref{eq:uz equation wavy} w.r.t $y$ and subjected to the boundary conditions~\eqref{eq:bcs for uz}, gives

\begin{equation}\label{eq:uz(y, z, t)}
    u_z(y,\, z,\, t)=\frac{G}{4}\qty(S_s^2-\qty(\alpha+y)^2+2S^2\ln\qty(\frac{\alpha+y}{S_s})) \quad \text{ for } y\in\qty[\zeta,\, \eta],
\end{equation}
where $G(z,\, t)\equiv \partial_z\kappa+Bo$.

\subsection{Interface equation}
Substitute~\eqref{eq:uz(y, z, t)} in the kinematic condition~\eqref{eq:kinematic bc wavy} and after evaluating the integral, we get
\begin{equation}
    S\pdv{\eta}{t}+\pdv{z}\qty{\frac{G}{16}\qty[2S^2\qty(S_s^2-S^2+2S^2\ln\qty(S/S_s))-\qty(S^2-S_s^2)^2]}=0,
\end{equation}
where $\partial_t\eta=\partial_tS$, hence, $S\,\partial_t\eta=\partial_t\qty(S^2)/2$ which leads to
\begin{equation}
    8\partial_t\qty(S^2)+\partial_z\,J=0,
\end{equation}
where the flux $J = \qty(\partial_z\kappa+Bo)\,M(S,\, S_s)$ with $M(S,\, S_s)$ as a mobility function given by 
\begin{equation}
    M(S,\, S_s)=2S^2\qty(S_s^2-S^2+2S^2\ln\qty(S/S_s))-\qty(S^2-S_s^2)^2,
\end{equation}
and $\kappa=\partial_{zz}S-S^{-1}$ is the curvature.

\backsection[Declaration of interests]{The authors report no conflict of interest.}
\backsection[Author contributions]{T.S. derived the model, performed the analysis and simulations, and wrote the paper. K.K.S. problem formulation. All authors reviewed the manuscript.
}
\backsection[Funding]{This research received no specific grant from any funding agency, commercial or not-for-profit sectors.}   
\backsection[Data availability statement]{The data underlying the figures are available from the corresponding author on reasonable request.}
  
\bibliographystyle{jfm}
\bibliography{jfm}

@article{craster2006viscous,
  title={On viscous beads flowing down a vertical fibre},
  author={Craster, RV and Matar, OK},
  journal={Journal of Fluid Mechanics},
  volume={553},
  pages={85--105},
  year={2006},
  doi={https://doi.org/10.1017/S0022112006008706},
  publisher={Cambridge University Press}
}

@article{wray2017reduced,
  title={Reduced models for thick liquid layers with inertia on highly curved substrates},
  author={Wray, Alexander W and Papageorgiou, Demetrios T and Matar, Omar K},
  journal={SIAM Journal on Applied Mathematics},
  volume={77},
  number={3},
  pages={881--904},
  year={2017},
  doi={https://doi.org/10.1137/16M1060686},
  publisher={SIAM}
}

@article{wierschem2003instability,
  title={Instability of a liquid film flowing down an inclined wavy plane},
  author={Wierschem, Andreas and Aksel, Nuri},
  journal={Physica D: Nonlinear Phenomena},
  volume={186},
  number={3-4},
  pages={221--237},
  year={2003},
  publisher={Elsevier}
}

@article{trifonov2007stability,
  title={Stability and nonlinear wavy regimes in downward film flows on a corrugated surface},
  author={Trifonov, Yu Ya},
  journal={Journal of applied mechanics and technical physics},
  volume={48},
  pages={91--100},
  year={2007},
  publisher={Springer}
}

@article{quere1999fluid,
  title={Fluid coating on a fiber},
  author={Qu{\'e}r{\'e}, David},
  journal={Annual review of fluid mechanics},
  volume={31},
  number={1},
  pages={347--384},
  year={1999},
  publisher={Annual Reviews 4139 El Camino Way, PO Box 10139, Palo Alto, CA 94303-0139, USA}
}

@article{trifonov1992steady,
  title={Steady-state traveling waves on the surface of a viscous liquid film falling down on vertical wires and tubes},
  author={Trifonov, Yu Ya},
  journal={AIChE journal},
  volume={38},
  number={6},
  pages={821--834},
  year={1992},
  publisher={Wiley Online Library}
}

@Article{ruyer2012wavy,
  title={Wavy regimes of film flow down a fiber},
  author={Ruyer-Quil, Christian and Kalliadasis, Serafim},
  journal={Physical Review E},
  volume={85},
  number={4},
  pages={046302},
  year={2012},
  publisher={APS}
}

@Article{sisoev2006film,
  title={Film flow down a fibre at moderate flow rates},
  author={Sisoev, Grigori M and Craster, Richard V and Matar, Omar K and Gerasimov, Sergei V},
  journal={Chemical engineering science},
  volume={61},
  number={22},
  pages={7279--7298},
  year={2006},
  publisher={Elsevier}
}

@article{solorio1987linear,
  title={Linear stability of a cylindrical falling film},
  author={Solorio, Francisco J and Sen, Mihir},
  journal={Journal of fluid mechanics},
  volume={183},
  pages={365--377},
  year={1987},
  publisher={Cambridge University Press}
}

@article{goren1962instability,
  title={The instability of an annular thread of fluid},
  author={Goren, Simon L},
  journal={Journal of Fluid Mechanics},
  volume={12},
  number={2},
  pages={309--319},
  year={1962},
  publisher={Cambridge University Press}
}

@article{kalliadasis1994drop,
  title={Drop formation during coating of vertical fibres},
  author={Kalliadasis, Serafim and Chang, Hsueh-Chia},
  journal={Journal of Fluid Mechanics},
  volume={261},
  pages={135--168},
  year={1994},
  publisher={Cambridge University Press}
}

@article{roy2002lubrication,
  title={A lubrication model of coating flows over a curved substrate in space},
  author={Roy, R Valery and Roberts, AJ and Simpson, ME},
  journal={Journal of Fluid Mechanics},
  volume={454},
  pages={235--261},
  year={2002},
  publisher={Cambridge University Press}
}

@article{roberts2006accurate,
  title={An accurate and comprehensive model of thin fluid flows with inertia on curved substrates},
  author={Roberts, AJ and Li, Zhenquan},
  journal={Journal of Fluid Mechanics},
  volume={553},
  pages={33--73},
  year={2006},
  publisher={Cambridge University Press}
}

@Article{ji2019dynamics,
  title={Dynamics of thin liquid films on vertical cylindrical fibres},
  author={Ji, H and Falcon, C and Sadeghpour, A and Zeng, Z and Ju, YS and Bertozzi, AL},
  journal={Journal of Fluid Mechanics},
  volume={865},
  pages={303--327},
  year={2019},
  publisher={Cambridge University Press}
}

@Article{grunig2012mass,
  title={Mass transfer characteristics of liquid films flowing down a vertical wire in a counter current gas flow},
  author={Gr{\"u}nig, J and Lyagin, E and Horn, S and Skale, T and Kraume, M},
  journal={Chemical engineering science},
  volume={69},
  number={1},
  pages={329--339},
  year={2012},
  publisher={Elsevier}
}

@Article{liu2021coating,
  title={Coating flows down a vertical fibre: towards the full Navier--Stokes problem},
  author={Liu, Rong and Ding, Zijing},
  journal={Journal of Fluid Mechanics},
  volume={914},
  year={2021},
  publisher={Cambridge University Press}
}

@Article{kliakhandler2001viscous,
  title={Viscous beads on vertical fibre},
  author={Kliakhandler, IL and Davis, Stephen H and Bankoff, SG},
  journal={Journal of Fluid Mechanics},
  volume={429},
  pages={381},
  year={2001},
  publisher={Cambridge University Press}
}

@Article{frenkel1992nonlinear,
  title={Nonlinear theory of strongly undulating thin films flowing down vertical cylinders},
  author={Frenkel, AL},
  journal={EPL (Europhysics Letters)},
  volume={18},
  number={7},
  pages={583},
  year={1992},
  publisher={IOP Publishing}
  }

@Article{duprat2007absolute,
  title={Absolute and convective instabilities of a viscous film flowing down a vertical fiber},
  author={Duprat, C and Ruyer-Quil, C and Kalliadasis, S and Giorgiutti-Dauphin{\'e}, F},
  journal={Physical review letters},
  volume={98},
  number={24},
  pages={244502},
  year={2007},
  publisher={APS}
}

@Article{ruyer2008modelling,
  title={Modelling film flows down a fibre},
  author={Ruyer-Quil, C and Treveleyan, P and Giorgiutti-Dauphin{\'e}, F and Duprat, C and Kalliadasis, S},
  journal={Journal of Fluid Mechanics},
  volume={603},
  pages={431--462},
  year={2008},
  publisher={Cambridge University Press}
}

@Article{duprat2009spatial,
  title={Spatial evolution of a film flowing down a fiber},
  author={Duprat, C and Ruyer-Quil, C and Giorgiutti-Dauphin{\'e}, F},
  journal={Physics of Fluids},
  volume={21},
  number={4},
  pages={042109},
  year={2009},
  publisher={American Institute of Physics}
}

@Article{gabbard2021asymmetric,
  title={Asymmetric instability in thin-film flow down a fiber},
  author={Gabbard, Chase T and Bostwick, Joshua B},
  journal={Physical Review Fluids},
  volume={6},
  number={3},
  pages={034005},
  year={2021},
  publisher={APS}
}

@Article{xie2021investigation,
  title={Investigation of flow dynamics of thin viscous films down differently shaped fibers},
  author={Xie, Qirui and Liu, Rong and Wang, Xun and Chen, Xue},
  journal={Applied Physics Letters},
  volume={119},
  number={20},
  pages={201601},
  year={2021},
  publisher={AIP Publishing LLC}
}

@Article{weinstein2004coating,
  title={Coating flows},
  author={Weinstein, Steven J and Ruschak, Kenneth J},
  journal={Annu. Rev. Fluid Mech.},
  volume={36},
  pages={29--53},
  year={2004},
  publisher={Annual Reviews}
}

@article{bonart2020stability,
  title={Stability of gravity-driven liquid films overflowing microstructures with sharp corners},
  author={Bonart, Henning and Rajes, Sangitha and Jung, Johannes and Repke, Jens-Uwe},
  journal={Physical Review Fluids},
  volume={5},
  number={9},
  pages={094001},
  year={2020},
  publisher={APS}
}

@PhdThesis{sadeghpour2020characteristics,
  title={Characteristics and Applications of Thin Liquid Films Flowing Down High-Curvature Surfaces},
  author={Sadeghpour, Abolfazl},
  year={2020},
  publisher={University of California, Los Angeles}
}

@Article{ji2020modelling,
  title={Modelling film flows down a fibre influenced by nozzle geometry},
  author={Ji, Hangjie and Sadeghpour, Abolfazl and Ju, Y Sungtaek and Bertozzi, Andrea L},
  journal={Journal of Fluid Mechanics},
  volume={901},
  year={2020},
  publisher={Cambridge University Press}
}

@Article{dietze2019effect,
  title={Effect of wall corrugations on scalar transfer to a wavy falling liquid film},
  author={Dietze, Georg F},
  journal={Journal of Fluid Mechanics},
  volume={859},
  pages={1098--1128},
  year={2019},
  publisher={Cambridge University Press}
}

@article{trifonov2017nonlinear,
  title={Nonlinear waves on a liquid film falling down an inclined corrugated surface},
  author={Trifonov, Yuri},
  journal={Physics of Fluids},
  volume={29},
  number={5},
  pages={054104},
  year={2017},
  publisher={AIP Publishing LLC}
}

@article{rivero2015nanomaterials,
  title={Nanomaterials for functional textiles and fibers},
  author={Rivero, Pedro J and Urrutia, Aitor and Goicoechea, Javier and Arregui, Francisco J},
  journal={Nanoscale research letters},
  volume={10},
  pages={1--22},
  year={2015},
  publisher={Springer}
}

@article{sarva2025dynamics,
  title={Dynamics of a liquid film flowing down a granular chain},
  author={Sarva, Kishor Kumar and Murthy, Tejas G and Tomar, Gaurav},
  journal={Journal of Fluid Mechanics},
  volume={1016},
  pages={A36},
  year={2025},
  publisher={Cambridge University Press}
}

\end{document}